\documentclass{article} 

\title{Quantum mechanical observers: a phase space approach}
\author{M. Dance}
\date{ }

\begin{document} 
\maketitle
 
\begin{abstract}
A quantum mechanical observer might be describable as having a reference system that is a superposition of classical inertial reference frames.  The present paper suggests a possible weighting function in such superpositions, determined by the product of the observer's wavefunction and the Fourier transform of the wavefunction at each point in phase space. This suggestion is made because each of these factors has a simple interpretation as a probability density amplitude.  Other forms for the weighting function may well be possible.
\end{abstract}

\section{Introduction} 
Heisenberg~\cite{heisenberg} noted that his uncertainty principle could apply to observers as well as to observed systems.  He noted that the uncertainties in an observer O1's position and momentum, as another observer O2 sees them, are given by:
\begin{equation}
\Delta x_{O1}^i   \Delta p_{O1}^i  \geq  \frac{\hbar}{2} .
\end{equation}
This means that O2 cannot know O1's position and velocity exactly at the same time.  By contrast, the Poincar\'e transformation of classical relativity requires that one observer knows the other's position and velocity exactly.  Heisenberg noted that his uncertainty relation was a problem for relativity, but postulated that any practical effect of observer indeterminacy could be eliminated by allowing the observer's mass to approach infinity.  However, a large observer mass warps spacetime.

These issues, and the dynamics of observers more generally, are often ignored in quantum mechanics. There has however been effort to construct quantum mechanical models of observers and their interactions with observed systems, in attempts over many decades to explain the measurement postulate of quantum mechanics, i.e. how wavefunctions of observed systems can collapse in finite time.  And by considering an observer's centre-of-mass backreaction during a measurement, Wigner~\cite{Wigner57} showed that there is a minimum observable distance uncertainty, or interval.  Such an interval features in theories involving e.g. noncommutative geometries, and in some of the present author's thoughts in e.g.~\cite{Dance0601} and~\cite{Dance0610}.

Quantum mechanical observers appear to be gaining ground. A nice review of work done on quantum reference frames is that of Bartlett et al~\cite{Bartlett06}, which has particular regard to precision optical experiments with transmitters and receivers at rest relative to each other.  Specific recent approaches to quantum mechanical observers in a relativistic setting include~\cite{WW06},~\cite{WW08} and~\cite{Dance0610}. These have discussed transformations between different observers, particularly the possibility of a quantum superposition of classical transformations. Other recent approaches to consistency of special relativity and quantum mechanics include e.g.~\cite{Kim&Noz06} and ~\cite{Kim&Noz08}. Another recent paper suggested that a phase space approach may be useful in this area. The paper~\cite{Dragoman0803} used a Wigner function approach in phase space to describe states of an observed system, under unitary transformations between reference frames, while the observers were not subject to a detailed description.

An aim of the present paper, as in~\cite{Dance0901}, is to look beyond the observer's centre of mass and understand how a fuller quantum description of the observer might impact on physics. The present paper characterises an observer in a pure quantum state as having a quantum superposition of classical reference frames.  The contribution of each classical frame to the superposition is postulated to be a product of the observer's wavefunction and the Fourier transform of the wavefunction, for the relevant point in phase space. There may well be other possible weighting formulae, e.g. involving quasiprobability functions.  The discussion will concern only observers with pure quantum mechanical states, but may be widened in future to include mixed states. Essentially, the suggested approach retains quantum mechanics while special relativity is modified in a way that does not touch its axioms. But because the wavefunction is a nonrelativistic entity, the suggested approach cannot unify quantum mechanics and relativity.
 
 
\section{Classical reference frames in special relativity}
 
We shall assume that there is a classical inertial reference frame OF.  Let us suppose that another classical inertial frame exists, with origin at $a$ and velocity $V$ relative to OF at $t_{OF} = 0$.  We will call this frame $F(a,V)$. We will assume that all such frames $F(a,V)$ can be used here, for all classically allowed values of $a$ and $V$.

What \emph{is} the classical inertial frame $F(a,V)$?  In special relativity, the frame has a notional set of clocks and rulers throughout spacetime.   In reality, there is no set of clocks and rulers filling spacetime.  The frame is largely notional, the main exception being its origin, where a real observer sits.  In practice, there may be a few other measuring devices in the frame, but we will view the reference frame as being really defined by the quantities that apply to the physical observer that sits at its origin.  That is, let our viewpoint be that $F(a,V)$ is defined by the relative position and velocity of its origin, relative to OF.  
 
 
\section{Quantum mechanical observer}
 
Continuing the above line of thought, how might we describe the reference system of a quantum mechanical observer O?  Instead of O having a specific position and velocity relative to OF at a given time $t_{OF}$, OF perceives O as having a wavefunction $\psi(x)$ at $t_{OF}=0$. (The present discussion concerns only an observer O "in" a pure quantum state.) To OF, O appears to have distributions of positions $x$ and momenta $p$.  
 
For any particular selection $(x,p)$ in phase space, let us postulate that we can ascribe to it a notional set of clocks and rulers filling spacetime, to give the frame $F(a,V(p,m_O))$ where $m_O$ is the rest mass of O, and $V$ is the classical velocity that corresponds to $p$ and $m_O$; $V = V(p,m_O)$.  Let us then suppose that OF can describe O as having each $F(a,V(p,m_O))$ in its set of frames, with some amplitude (density) $W(a,p)$ determined by $\psi(x)$.  A guess could be $W(a,p) = \psi(x) \phi(p)$, where $\phi(p)$ is the Fourier transform of $\psi(x)$.  More generally, one could perhaps use some other function, e.g. a quasiprobability distribution such as the Wigner function.  The suggestion $\psi(x) \phi(p)$ is made because each of these factors has a simple interpretation as a probability (density) amplitude.  

Now we will postulate that we can add up all of these $F(a,V)$ classical inertial frames, with their respective amplitude weightings, to give the observer O's reference system, as far as OF knows it. We will call this reference system simply "O".  Then:
\begin{equation}
\label{O} 
O = \int da dp  W(a,p) F(a,V(p,m_O)) .
\end{equation} 
 
It may be worth considering what this integral means, and how to carry it out.  In particular, one could ask whether the integral over $a$ should be over all four components of $a$ including the time component, or whether it might be preferable to integrate only over the spatial components of $a$.  If the first approach is taken, including O's wavefunction throughout all of spacetime, O's reference system is a constant system that superposes all contributing classical inertial frames.  If the second approach is taken, one sees how O's reference system changes with time, which seems to be a more natural approach.  In that approach, the time variable in O(t) is OF's time coordinate. Both approaches may be equivalent; the second is a time slicing of the first. (In future, one could perhaps find $O(t_O)$ by mapping $(t_O, x_O)$ back to corresponding superpositions of coordinates in OF and superposing the results of Equation~\ref{O} as appropriate.) In this paper, $x$ will sometimes be used to mean spatial components only, and sometimes will include all four components.  The context should make it clear which is meant in any instance.

\section{Transformations between reference systems}
 
The transformation from OF to O is straightforward.  It is a superposition of classical Poincar\'e transformations.  This can be seen by noting that each inertial frame $F(a,V)$ in Equation~\ref{O} above can be obtained from OF by a Poincar\'e transformation, which we shall call $Pe(a,V)$:
\begin{equation} 
O = \int da dp  W(a,p) F(a,V(p,m_O))  =  \int da dp  W(a,p) Pe(a,V(p,m_O)) OF.
\end{equation} 
 
The transformation between two quantum mechanical observers O1 and O2 is less trivial. Their reference systems are given, relative to OF, by:
\begin{equation} 
O1 = \int da_1 dp_1  W(a_1,p_1) F(a_1,V_1(p_1, m_{O_1})) 
\end{equation}
and
\begin{equation} 
O2 = \int da_2 dp_2  W(a_2, p_2) F(a_2,V_2(p_2, m_{O_2}) )
\end{equation}
It would be desirable to prove that the transformation between O1 and O2 is unitary. This is left for future work and will not be discussed further in this paper. 

\section{Example} 
\label{example1}
To illustrate the above ideas, let us now work out how Equation~\ref{O} looks when O's wavefunction is a plane wave in OF's frame:
\begin{equation} 
\psi_O (x) = \frac{1}{\sqrt{2\pi\hbar}} e^{ik.x}
\end{equation}
where $k.x = k_\mu x^\mu = -k^0 x^0 + k^1 x^1 + k^2 x^2 + k^3 x^3$.  The Fourier transform of $\psi_O(x)$ is $\phi (p) = \delta (p - \hbar k)$.  Then
\begin{equation}
W(a,p) = \frac{1}{\sqrt{2\pi\hbar}} e^{ik.a} \delta (p - \hbar k)       
\end{equation}
and
\begin{equation}
O = \int da dp \frac{1}{\sqrt{2\pi\hbar}} e^{ik.a} \delta (p - \hbar k)  F(a,V(p,m_O))
\end{equation}
which simplifies to
\begin{equation}
O = \int da \frac{1}{\sqrt{2\pi\hbar}} e^{ik.a}  F(a,V(\hbar k, m_O))
\end{equation}
This is a superposition of classical inertial frames, each contributing the same magnitude for all $a$ and weighted by a varying phase factor. Physically, the equal magnitudes might be expected because OF is equally likely to find O anywhere in space(time).
 

\section{A system observed by OF and O}
 
So far we have considered only the reference frames/systems of OF and O.  We will now consider how OF and O would describe an observed system, sometimes called in the literature a "system under observation" or SUO.
 
Suppose OF describes the SUO by a scalar wavefunction $\chi(x,t)$ (or equivalently $\chi (x)$ with $x$ in 4-dimensional spacetime). How does O describe the SUO?  (In standard Copenhagen quantum mechanics, perhaps all we can say is how OF believes O will describe the SUO!)

Consider the 4-vector $x$, which OF views as a single spacetime point.  By contrast, O views the point as relative to a reference system which is a superposition of classical inertial frames.  We postulate that O (effectively) evaluates the coordinates of $x$ in each of those reference frames, to an extent corresponding to the frame amplitude.  In the frame $F(a,V(p, m_O))$, the components of $x$ are transformed to the new components ${x^\prime }$ with
\begin{equation} 
x^\prime = Pe(x, a, V(p, m_O)) 
\end{equation}
where the right hand side means the coordinates of $x$ in OF's frame are Poincar\'e-transformed to those in frame $F(a,V(p, m_O))$.  (We retain the property that the point $x$ is invariant, while the components of $x$ change between frames.)

Then we postulate that O describes the SUO by the wavefunction $\chi^{\prime}(x)$, where
\begin{equation} 
\label{chiprime}
\chi^\prime (x) = \int da dp db \delta( x  -  Pe(  b, a, V(p, m_O)  )    ) W(a,p) \chi (b)
\end{equation}
Equation~\ref{chiprime} sums over all combinations of coordinates (of $b$) in OF and Poincar\'e transformations (determined by $a$ and $V$) which end up with the coordinates $x$ in O.  The expression $Pe(b,a,p)$ is used here to denote the result when the transformation $Pe(a,V(p, m_O))$ is applied to the coordinates of $b$.    Each transformation contributes with the corresponding amplitude $W(a,p)$. The integrals over $a$, $b$ and $p$ are over all four components of each variable. Some thought could be given in future to the integration limits that might be appropriate.

Equation~\ref{chiprime} can be simplified to: 
\begin{equation}
\label{chipri}
\chi^{\prime}(x) = \int da dp W(a,p) \chi ( Pe( x, -\Lambda^{-1}(V) a, -V ) )
\end{equation}
where $V = V(p, m_O)$.  
 
\section{Example}
Let us now return to the example of Section~\ref{example1}, in which O is a plane wave, with 
\begin{equation}
\psi_O(x) = \frac{1}{\sqrt{2\pi \hbar}} e^{ik.x}
\end{equation} 
and 
\begin{equation}
W(a,p) = \frac{1}{\sqrt{2\pi\hbar}} e^{ik.a} \delta (p - \hbar k) . 
\end{equation}
Inserting these expressions into Equation~\ref{chipri}, we have
\begin{equation}
\label{ex}
\chi^{\prime}(x) = \frac{1}{\sqrt{2\pi \hbar}}  \int da e^{ik.a} \chi( Pe( x, -\Lambda^{-1}(V) a, -V ) )
\end{equation}
where $V = V(\hbar k, m_O)$.

Let us now take, as an hypothetical example, the SUO to be described by OF as a plane wave, with the same wavefunction as O, and let us integrate over all four components of $a$ with infinite limits. Then it turns out from Equation~\ref{ex} that as far as OF knows, O will describe the SUO by the wavefunction: 
\begin{equation}
\chi^{\prime}(x) = \frac{1}{2\pi \hbar} e^{ik.\Lambda(-V).x}
\end{equation} 
 
 
\section{Discussion}
 
The formalism suggested in the present paper uses a simple formula for $W(a,p)$.  Other possibilities may include e.g. a suitable quasiprobability function.

The present paper has used the standard Copenhagen interpretation of quantum mechanics, which views $\psi_O(x)$ as representing the knowledge that another observer OF has about the observer O, rather than the state of the observer O itself.  However, other interpretations of quantum mechanics view $\psi_O(x)$ as representing reality, i.e. a real state of O.  We have not entered this epistemology/ontology debate here. It may be that $\psi_O(x)$ really \emph{is} the state of O, and if so, Equation~\ref{O} could simply \emph{be} O's reference system.  In any case, these issues of interpretation could perhaps give rise to interesting work in the area of quantum reference frames. 

While the use of quantum mechanical reference systems (or some other theory entirely) might be required to fully unite quantum mechanics and relativity, the formalism suggested here uses nonrelativistic quantum mechanics.  This appears to inherently limit the scope of applicability of the suggestions made in the present paper.  However, the suggestions could perhaps serve as a starting point for further development. For example, in quantum field theory, one might write something similar to
\begin{equation}
\phi^{\prime}(x^\prime) = \int d^4 x d^4 a d^4 p W(a,p) \phi(x) \delta(x - Pe(x^\prime;-\Lambda^{-1}(V)a, -V))
\end{equation}
for a scalar field $\phi$ as seen by OF, and seen as $\phi^\prime$ by O.  The amplitude factor $W(a,p)$ would be determined by a field that describes the observer O. 

 
 
\end{document}